\documentclass[amsmath,amssymb,aps,prl,reprint, showpacs, superscriptaddress]{revtex4-2}
\usepackage{graphicx}
\usepackage{amsthm}
\usepackage{amsmath}
\usepackage{amssymb}
\usepackage{dsfont}
\usepackage{bm}
\usepackage{epstopdf}
\usepackage{hyperref}
\usepackage{xcolor}

\begin{document} 
\title[]{Finite-size scaling at the edge of disorder in a time-delay Vicsek model}
\author{Viktor Holubec}
\email{viktor.holubec@mff.cuni.cz}
\affiliation{ 
 Charles University,  
 Faculty of Mathematics and Physics, 
 Department of Macromolecular Physics, 
 V Hole{\v s}ovi{\v c}k{\' a}ch 2, 
 CZ-180~00~Praha, Czech Republic 
}
\author{Daniel Geiss}
\affiliation{ 
Institut f{\"u}r Theoretische Physik, 
Universit{\"a}t Leipzig, 
Postfach 100 920, D-04009 Leipzig, Germany
}
\affiliation{ 
Max Planck Institute for Mathematics in the Sciences, 
D-04103 Leipzig, Germany
}
\author{Sarah A.M. Loos}
\affiliation{ 
Institut f{\"u}r Theoretische Physik, 
Universit{\"a}t Leipzig, 
Postfach 100 920, D-04009 Leipzig, Germany
}
\author{Klaus Kroy}
\affiliation{ 
Institut f{\"u}r Theoretische Physik, 
Universit{\"a}t Leipzig, 
Postfach 100 920, D-04009 Leipzig, Germany
}
\author{Frank Cichos}
\affiliation{Peter Debye Institute for Soft Matter Physics, Universit\"at Leipzig, 04103 Leipzig, Germany.}
\date{\today} 


\begin{abstract} 
Living many-body systems often exhibit scale-free collective behavior reminiscent of thermal critical phenomena. But their mutual interactions are inevitably retarded due to information processing and delayed actuation. We numerically investigate the consequences for the finite-size scaling in the Vicsek model of motile active matter. A growing delay time initially facilitates but ultimately impedes collective ordering and turns the dynamical scaling from diffusive to ballistic. It provides an alternative explanation of swarm traits previously attributed to inertia.
\end{abstract}

\maketitle  


Interacting assemblies of active elements ranging from neural networks in the brain to forest fires and bird flocks can exhibit scale-free behavior~\cite{mora2011biological,barabasi2003,munoz2018colloquium,Cavagna2018a}. This might be indicative of an underlying powerful physical ordering principle overwriting their inherent complexity.  
Finite-size scaling theory~\cite{plischke2006} associates such behavior with a correlation length exceeding the system size and conjectured to arise from a mechanism called self-organized critically~\cite{Bak1988}. It is indeed an appealing idea that simple interaction rules, such as an innate tendency of individual active elements to replicate the action of their neighbors, can drive a non-equilibrium ensemble towards criticality. Yet, it does not generally seem to apply to both natural systems~\cite{Clauset2009,Jhawar2020} and their models~\cite{Pruessner2002,Palmieri2020,Chate2021}.

Studying the emergent finite-size scaling in natural assemblies is thus vital for their prospective modeling in the spirit of non-equilibrium many-body systems, as successful models should be required to reproduce the observed scaling~\cite{Palmieri2020} and correlations~\cite{Cavagna2019}. In this vein, the inertia spin model~\cite{Cavagna2019} was proposed to overcome known deficiencies of the classical Vicsek model (VM) \cite{vicsek1995novel} in comparison with empirical data for natural swarms and flocks. Inspired by observations of birds and insects~\cite{Cavagna2018a}, which cannot turn instantaneously, it adds inertia to the navigation rules for the individual motile spins. The dynamical scaling 
and time correlation functions found in natural swarms~\cite{Cavagna2017} of moderate size (dynamical exponent $z\approx 1.1$) could thereby be reconciled with the classical VM prediction ($z=2$)~\cite{Cavagna2018a}.

However, it seems worthwhile to point out that, for motile ensembles, physical inertia can have quite similar effects as delayed reactions  due to finite speeds of information transfer and processing, and actuation~\cite{nagy2010hierarchical,Attanasi2014,geiss2019brownian}. The latter are indeed ubiquitously found both in nature and in engineering practice, from insects to birds and various robotic systems~\cite{vasarhelyi2014outdoor,Viragh2014,Vasarhelyieaat2018}, and also thought to cause traffic jams~\cite{Davis2003}. Recent experiments~\cite{khadka2018active,Munos2021} with feedback-driven artificial microswimmers~\cite{Franzl2021} have moreover established the role of delayed interactions also in the microscopic world of active Brownian particles.
Beyond oscillatory behavior, which is also a common trait of inertial motion, time-delayed interactions can give rise to multistability, instabilities, and even chaos~\cite{forgoston2008delay,piwowarczyk2019influence,geiss2019brownian,loos2019heat}. Conversely,  intermediate time delays may facilitate clustering compared to the classical VM~\cite{piwowarczyk2019influence} and flocking in the Cucker--Smale model~\cite{Erban2016}. And recent indications that delay-dependent optimizations play a role in artificial microswimmer assemblies~\cite{Munos2021} seem reminiscent of the optimum run-and-tumble times of bacteria~\cite{Romanczuk2015,Diz-Munoz2016} or the improved localisation achieved with feedback cooling~\cite{Bushev2006,Goldwater2019} or feedback-driving of robots~\cite{mijalkov2016engineering}.

\begin{figure}
\includegraphics[width=0.6\columnwidth]{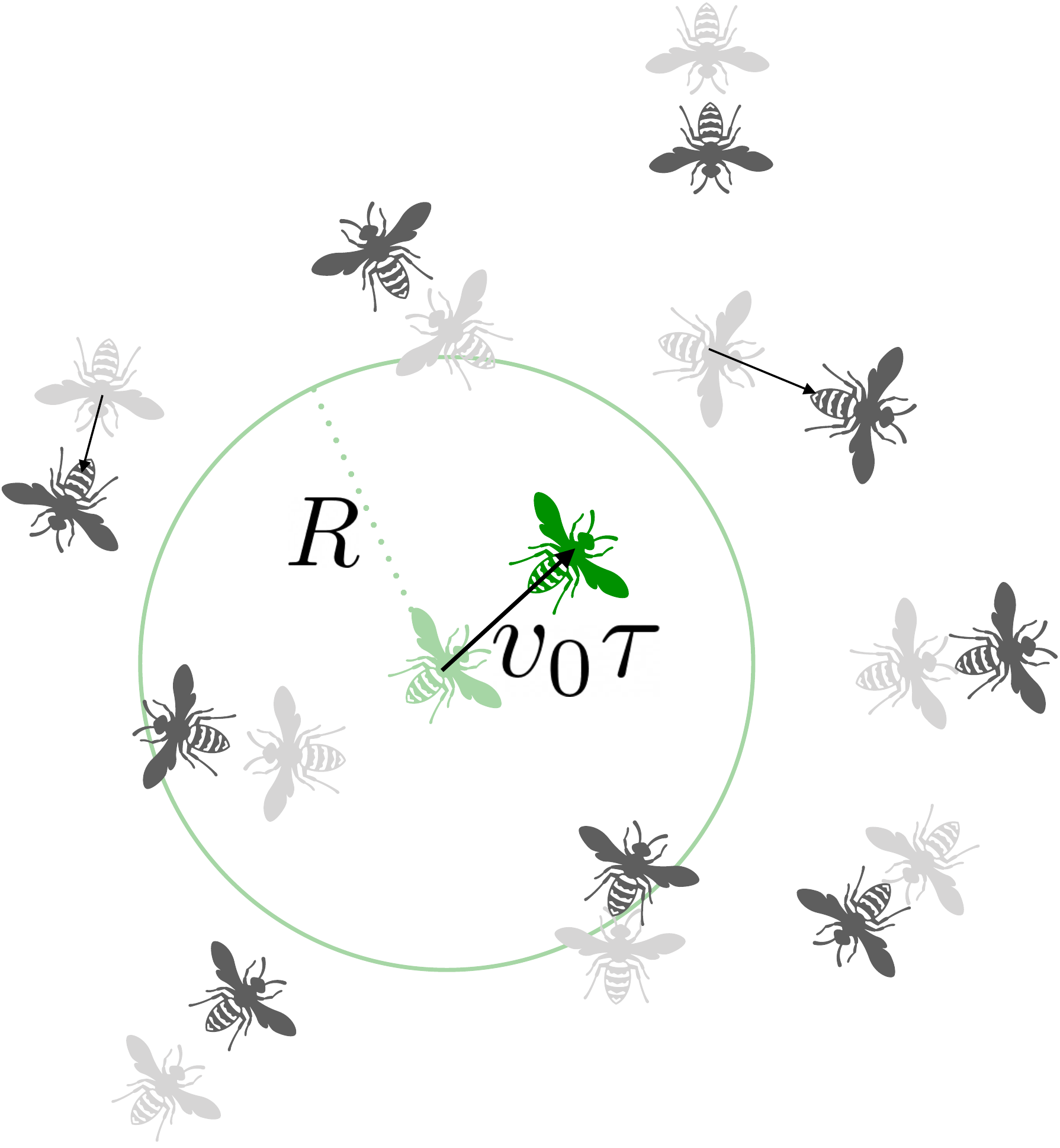}
	\caption{Sketch of the delayed Vicsek model defined by Eqs.~\eqref{eq:vtdisc} and \eqref{eq:rtdisc}. The conventional alignment rule with interaction range $R$ is executed with a time delay $\tau$.} 
	\label{fig:sketch}	
\end{figure}

In this Letter, we demonstrate that by introducing retarded reactions in the classical VM~\cite{vicsek1995novel}, we can obtain the same finite-size scaling and time correlations near the ordering transition as previously obtained with the inertia spin model~\cite{Cavagna2018a}. This suggests that similar scaling and correlations as observed for natural swarms can be expected for a wealth of systems with time-delayed interactions, including over-damped Brownian particle assemblies. We can also corroborate the observation that increasing delay times may have a non-monotonic effect on the stability of coherent collective motion \cite{piwowarczyk2019influence}.

\textit{Model:} The (classical) VM~\cite{vicsek1995novel} arguably is the simplest model for motile active-particle assemblies, ranging from bacteria to birds, and a central paradigm in the field motile active matter~\cite{bechinger2016active}. In each discrete time step, all particles advance with the same constant speed $v_0$. And they instantaneously adapt their orientations to the previous average orientation of their neighbors within an interaction sphere of radius $R$, up to some random error contributed by a local noise term. The orientation is thus clearly an overdamped variable, as it bears no inertia.

In the delay VM, depicted in Fig.~\ref{fig:sketch}, particle $i$ adapts at time $t$ to the mean orientation of all particles that had distance less than $R$ from its previous position, at time $t -1 -\tau$, with an integer-valued time delay $\tau \ge 0$. The discrete time step and the interaction radius $R=1$ serve as units of time and length, respectively. 
The dynamics of the standard VM is recovered for $\tau = 0$. The velocity $\mathbf{v}_i$ and position $\mathbf{r}_i$  of particle $i$ in three spatial dimensions (3D) thus obey the set of equations~\cite{vicsek1995novel}
\begin{align}
    \mathbf{v}_i(t+1) &= v_0 \mathcal{R}_\alpha \Theta\!\left[
    \mathbf{v}_i(t) + \sum_{j} n_{ij}(t-\tau) \mathbf{v}_j(t - \tau) 
    \right],
    \label{eq:vtdisc}\\
    \mathbf{r}_i(t+1) &= \mathbf{r}_i(t) + \mathbf{v}_i(t+1).
    \label{eq:rtdisc}
\end{align}
The noise operator $\mathcal{R}_\alpha X$ randomly rotates its argument $X$ within a uniformly distributed solid angle $4\pi\alpha$ centered around $X$, and $\Theta(\mathbf{v}) \equiv \mathbf{v}/|\mathbf{v}|$ normalizes its argument. 
We assume geometric interactions corresponding to the connectivity matrix elements $n_{ij}(t) = 1$ for  $i\neq j$ if $r_{ij}(t) = |\mathbf{r}_i(t) - \mathbf{r}_j(t)| < R$, and $n_{ij}(t) = 0$ otherwise.

We simulated the delay VM with fixed speed $v_0 = 0.05$ and noise strength $\alpha = 0.45$ inside a cube with size $L^3$ and periodic boundary conditions for six values of the particle number $N = 2^n$, $n = 6,\dots,11$. In this setting, we repeated the analysis performed in Refs.~\cite{attanasi2014finite, Cavagna2017} for the static and dynamic scaling and the correlation functions of the standard VM, for delay times $\tau = 0,\dots,20$. As control parameter, we used the average nearest-neighbor distance $r_\mathrm{1}$ between the individual particles, which we changed by varying $L$. Here, we present the main results from the simulations. Further data, technical details, and some analytical discussion can be found in the Supplemental Material~\cite{SI}. 

The central object for our data analysis is the Fourier transformed spatio-temporal correlation function (CF) 
\begin{equation}
C(k,t) =
\left<
\frac{1}{N} \sum_{i,j}^N
\frac{\sin[k \,r_{ij}(t,t_0)]}{k r_{ij}(t,t_0)}
\delta \hat{\mathbf{v}}_i(t_0) \cdot \delta \hat{\mathbf{v}}_j(t_0+t)
\right>
\label{eq:corr}
\end{equation}
of the normalized velocity fluctuations ~\cite{SI,Cavagna2017, Cavagna2018a} \begin{equation}
\label{eq:fluc}
	\delta\hat{\mathbf{v}}_i = \frac{\delta\mathbf{v}_i}{\sqrt{N^{-1} \sum_k \delta\mathbf{v}_k \cdot \delta\mathbf{v}_k}},
\end{equation}
where $\delta\mathbf{v}_i = \mathbf{v}_i - \sum_k \mathbf{v}_k/N$  is the deviation of the velocity of particle $i$ from the average velocity, and $r_{ij}(t_0,t) = |\mathbf{r}_i(t_0) - \mathbf{r}_j(t)|$ the distance between particles $i$ and $j$ at times $t$ and $t_0<t$. The average $\left<\dots\right>$ is taken over  $t_0$~\cite{SI}.

\textit{Static scaling:} At $t=0$, $C(k,0)$ exhibits a global maximum at $k=k^\star \sim 1/\xi$, where $\xi$ corresponds to the correlation length. Assuming proportionality between fluctuation and response, this value of the CF is interpreted as a susceptibility $\chi \equiv C(k^*,0)$  \cite{SI,Cavagna2017, Cavagna2018a}.

\begin{figure*}
	\centering
\includegraphics[width=1.0\linewidth]{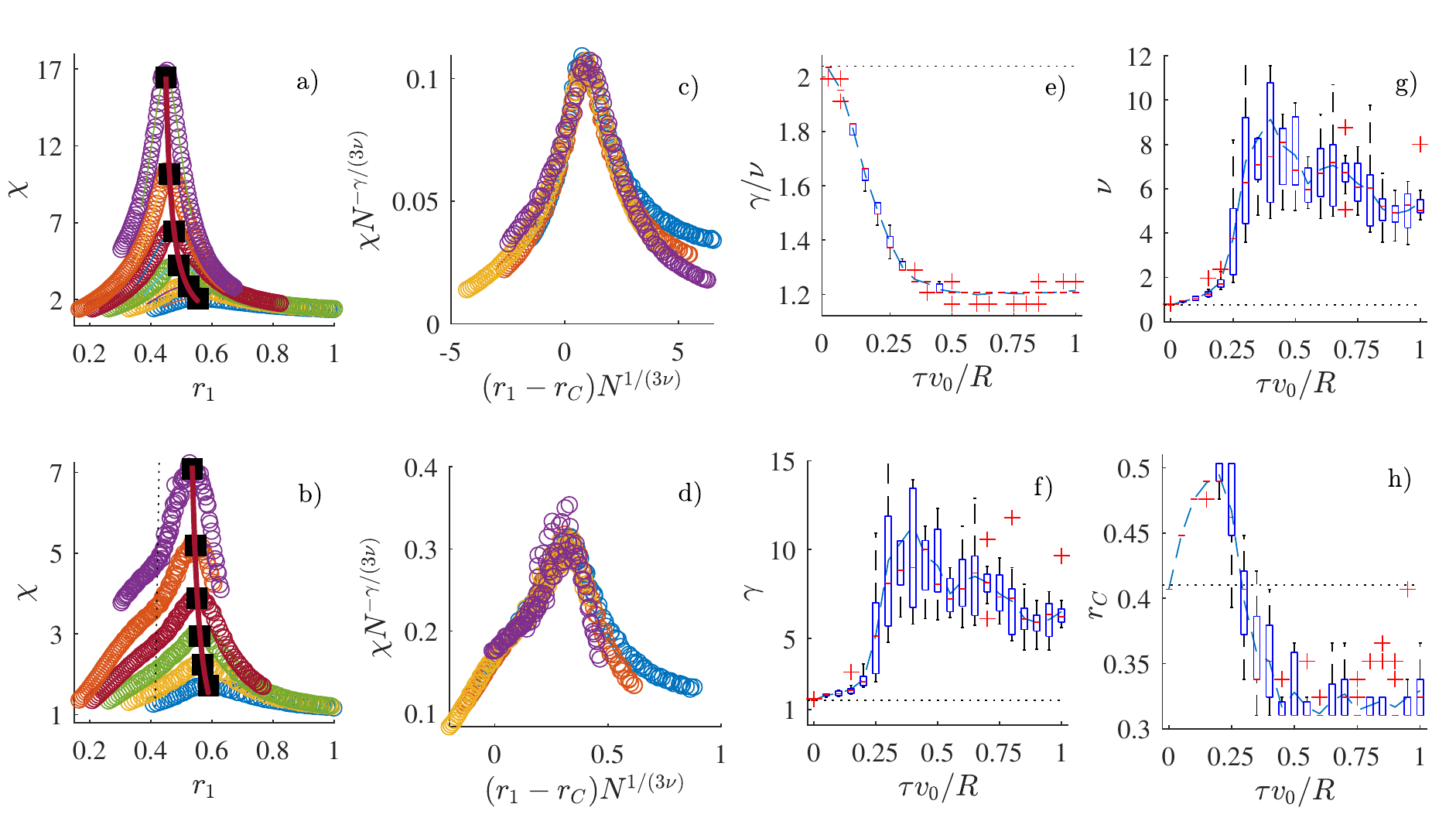}
	\caption{Static finite-size scaling in the 3D delay VM. Each simulation departed from a random initial state and was evolved for a transient period of 1500 time-steps before measurements started. Panels a) and b) show the susceptibilities $\chi \equiv \max_k C(k,0)$ averaged over 11 trajectories of $10^4$ time-steps
	for $N=64,128,256,512,1024,2048$ particles (from bottom to top) and delay times $\tau = 0$ [a), standard VM] and $\tau = 15$ [b), $\tau v_0/R=0.75$], respectively, over the mean nearest neighbour distance $r_1$. Panels c) and d) illustrate the data collapse achieved for $N\geq 256$, e)-h) show box-plots~\cite{boxplot} of 
	the exponent ratio $\gamma/\nu$, the individual exponents, and the extrapolated critical parameter $r_C$ for $N\to\infty$, respectively, all for delay times $\tau$ from 0 to 20. Broken dashed lines mark averages over the 11 realizations. The horizontal dotted lines depict the values $\nu \approx 0.75$, $\gamma \approx 1.53$, and $r_C \approx 0.41$ obtained for $\tau =0$, where the model reduces to the standard VM, consistent with the data in Ref.~\cite{attanasi2014finite}.
	In a) and b) black squares mark susceptibility maxima $\chi(r_1^*)$ and solid lines are computed using the finite-size scaling relation $\chi \sim (r_\mathrm{1}^* - r_{C})^{-\gamma}$ with parameters from e)-h). The vertical dotted line in b) marks the abrupt changes in the slope of $\chi$.}
	\label{fig:static}
\end{figure*}

For given delay time $\tau$ and particle number $N$, the susceptibility $\chi$ exhibits a maximum $\chi^* = \chi^*(\tau, N)$ as function of the nearest neighbour distance at $r_1^* = r_1^*(\tau,N)$. The system is found to be ordered (large average velocity) for $r_1 < r_1^*$ and disordered (small average velocity) otherwise. For given $N$, the susceptibility $\chi^*$ at the transition  decreases monotonically with growing $\tau$ and eventually saturates, see Figs.~\ref{fig:static}a and b and the supplemental Figs.~S1 and S2~\cite{SI}. The equal-time orientation correlations are thus generally reduced for retarded as opposed to instantaneous interactions, which suggests that the sensitivity to external perturbations decreases accordingly. For sufficiently large $\tau$ and $N$, the derivative of the susceptibility with respect to $r_1$ abruptly increases at some $r_1 < r_1^*$, see the vertical dotted line at $r_1 \approx 0.4$ in Fig.~\ref{fig:static}b. No such kink is observed for small $\tau$~\cite{SI}.

For given $\tau$ and large enough $N$, the susceptibility in the vicinity of the ordering transition~\cite{attanasi2014finite} exhibits finite-size scaling according to~\cite{fisher1972scaling}
\begin{align} 
	r_\mathrm{1}^* &\sim r_{C} + N^{-1/(3\nu)},  \label{eq: scaling1}\\
	\chi &\sim N^{\gamma/(3\nu)} \label{eq: scaling2}.
\end{align} 
In other words, for any given $\tau$, the limiting location  $r_C = r_\mathrm{1}^*(\tau,\infty)$ of the transition for large (infinite) particle numbers and the critical exponents $\gamma$ and $\nu$ of the susceptibility $\chi \sim (r_1^* - r_{C})^{-\gamma}$ and the correlation length $\xi \sim (r_1^* - r_{C})^{-\nu}$, respectively, can all be extrapolated from a data collapse of the susceptibilities for different $N$. The procedure is illustrated
 in Figs.~\ref{fig:static}c and d. The resulting exponents and $r_C$ in Figs.~\ref{fig:static}e-h exhibit strong dependencies on $\tau$, which saturate as $\tau v_0/R \approx 1/2$, as also suggested by an analytical argument~\cite{SI}.

Since the static critical exponents in the standard VM are known to depend strongly on the density, speed and interaction radius~\cite{cambui2016critical}, their absolute values are of limited interest. Rather, their trends and dependencies are revealing. The critical nearest-neighbor distance $r_C$ in Fig.~\ref{fig:static}h exhibits a pronounced maximum at $\tau v_0/R \approx 0.2$, indicating that a system with an intermediate delay time favors order already at lower densities as compared to the system without delay. This somewhat counter-intuitive result is in agreement with findings of Refs.~\cite{mijalkov2016engineering,piwowarczyk2019influence} that intermediate delays stabilize collective motion. For larger delay times,
the critical nearest-neighbor distance $r_C$ drops sharply to a value below that for the standard VM.

The exponents $\nu$ and $\gamma$ also display local maxima, but at somewhat larger $\tau$. Their saturation values are much higher than the respective critical exponents in all known universality classes, including the standard VM. To attribute the observed finite-size scaling to a critical point in the infinite-size limit according to the conventional scaling hypothesis \cite{halperin1969scaling} would require an extraordinarily sharp divergence of the correlation length and the susceptibility at criticality, approached extraordinarily slowly with increasing particle number $N$, see the solid lines in Figs.~\ref{fig:static}a and b. 

\begin{figure*}
	\centering
	\includegraphics[width=1.0\linewidth]{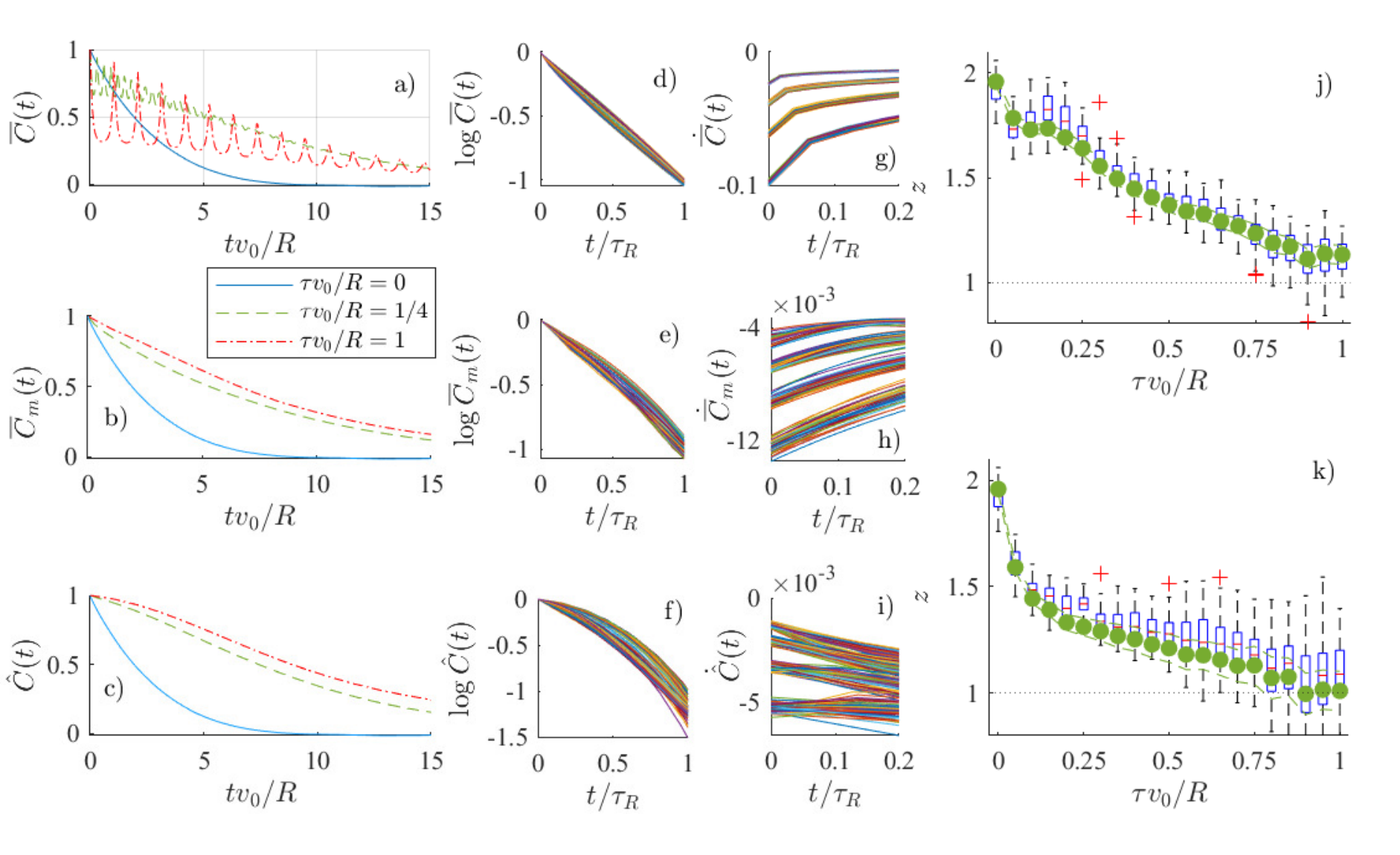}
	\caption{Dynamical scaling of the orientational correlations at the susceptibility maximum displayed in Fig.~\ref{fig:static}. a) Normalized time-correlation functions $\overline{C}(t)$ for delays $\tau=0,5,20$ and $N=2048$. b) The upper envelopes $\overline{C}_m$ of the curves in a). c) The correlation functions $\hat{C}(t)$ calculated from trajectories (under)sampled with frequency $1/(\tau + 1)$ for $N=2048$. d) - f) The normalized CFs $\overline{C}$, $\overline{C}_m$ and $\hat{C}$ collapse upon measuring time in units of the relaxation time $\tau_R$ for $\tau=0$ (d) and $\tau=20$ (e) and (f) and system sizes $N= 2^n$, $n=8,\dots,11$  ($50$ simulation runs for each system size); g) -- i) the corresponding slopes. Box-plots~\cite{boxplot}  in j), k) show the dynamical  exponent $z$ obtained from  collapsing the CFs $\overline{C}_m$, $\hat{C}$ by rescaling time as $t \xi^z$. The green circles are from linear fits to $\log \tau_R(N) = - z \log k^*(N)$ for all 50 data sets with 95\% confidence intervals of the fits (dashed).}
	\label{fig:dynamic}
\end{figure*}

\textit{Time correlation functions:} The time dependence of the CFs~\eqref{eq:corr} for the delay VM at the transition, 
quantifying the temporal loss of orientational correlations~\cite{Cavagna2017,Cavagna2018a}, is strongly influenced by the delay. Figure~\ref{fig:dynamic}a shows that the normalized CFs $\overline{C}(t) \equiv C(k^*,t)/C(k^*,0)$ acquire oscillations with period ($\tau + 1$) and an amplitude increasing with $\tau$. They reveal the transmission of orientational correlations over discrete time steps $\tau + 1$ and can be understood analytically by a spin-wave theory that accounts for the delay~\cite{SI}.  

In Fig.~\ref{fig:dynamic}d, we show that logarithms of CFs for $N = 2^n$, $n=8,\dots,11$ and $\tau = 0$ collapse onto the master curve $-t/\tau_R$ upon rescaling time by the relaxation times $\tau_R$ obtained from Eq.~(S14)~\cite{SI}. Figure~\ref{fig:dynamic}g shows the corresponding increasingly negative time derivatives for $t \to 0$,
indicating the exponential loss of correlation in the
standard VM~\cite{Cavagna2017}.

Due to the delay-induced superimposed oscillations, the initial slope of the CFs always steepens with increasing delay time $\tau$. In contrast, the \emph{overall} decay flattens with  $\tau$, as revealed by the upper envelopes $\overline{C}_m(t)$ of $\overline{C}(t)$, in Fig.~\ref{fig:dynamic}b. As intuitively expected, the delayed interactions thus tend to increase the memory in the VM. The data collapse of $\overline{C}_m(t/\tau_R)$ in Fig.~\ref{fig:dynamic} confirms the non-exponential relaxation, while its slope in Fig.~\ref{fig:dynamic}h still increases for $t \to 0$ (see Sec.~S5~\cite{SI} for an approximate analytical result). 

However, we note that the sampling rates used in practical measurements may not always be sufficient to resolve delay induced oscillations~\cite{SI}, which might moreover have a tendency to be washed out by a natural dispersion of the delay times. To account for such undersampling, Fig.~\ref{fig:dynamic}c shows normalized CFs $\hat{C}(t) = C_{\rm C}(k^*,t)/C_{\rm C}(k^*,0)$ calculated from particle positions that were (under-)sampled with frequency $1/(\tau + 1)$, i.e. we calculated the corresponding velocities as $\mathbf{v}_i(t) = [\mathbf{r}_i(t) - \mathbf{r}_i(t-\tau - 1)]/(\tau + 1)$. In Fig.~S9~\cite{SI} we show that the undersampled CFs are independent of the sampling rate as long as it is comparable to or smaller than $1/\tau$. The resulting CFs are shown in Fig.~\ref{fig:dynamic}c. The exponential initial decay for vanishing delay time $\tau = 0$ is seen to increasingly flatten with growing $\tau$, as also corroborated by the data collapse in Fig.~\ref{fig:dynamic}f. 
For $v_0 \tau \gtrsim 0.5$, the absolute slope $|\dot{\overline{C}}(t)|$ starts to decrease for $t\to 0$ as shown in Fig.~\ref{fig:dynamic}i and Figs.~S6 and S8~\cite{SI}. 

The undersampled delay VM yields qualitatively the same relaxation of orientational correlations as observed for natural swarms~\cite{Cavagna2017}. It thus provides an alternative explanation for the data, which where so far interpreted within the inertia spin model~\cite{Cavagna2018} --- the version of the VM in which the particle reorientation exhibits inertial dynamics, which amounts to another form of time-delayed alignment interactions. The dynamics induced by a discrete time delay appears to be more prone to developing oscillatory patterns, as the ones illustrated in Fig.~\ref{fig:dynamic}a, though. 

\textit{Dynamical scaling:}
The dynamical scaling hypothesis~\cite{halperin1969scaling} states that the relaxation time $\tau_R$ diverges with the correlation length as $\tau_R \sim \xi^z \sim (k^*)^{-z}$, at a critical point. Directly fitting the relation 
\begin{equation}
\log \tau_R(N) = - z \log k^*(N)    
\end{equation}
for the relaxation times of $\overline{C}_m$ and $\hat{C}$ as functions of time delay $\tau$ yield the dynamical exponent $z$ as depicted by circles in Figs.~\ref{fig:dynamic}j and k, respectively. The figures also show box-plots~\cite{boxplot} resulting from the best data collapse of CFs $\overline{C}_m$ and $\hat{C}$ using $(k^*)^{-z}$ with $z$ as a free parameter in place of the relaxation time $\tau_R$, itself. 
The exponents $z$ obtained from these two approaches nicely agree for each CF. While the values for $z$ obtained from the two alternative CFs differ, they exhibit the same robust trend: a crossover from a ``diffusive'' to a ``ballistic'' signature, i.e., from $z\approx 2$ for $\tau = 0$ to about $z\approx 1.1$ for $\tau v_0/R \approx 1$. The saturation of $z(\tau)$ for large $\tau$ follows from that of $\xi(\tau)$ (implied by the saturation of the finite-size scaling parameters in Fig.~\ref{fig:static}) and $\tau_R(\tau)$ (Fig.~S11~\cite{SI}).
These results can therefore reconcile the standard VM predictions with the observations for natural swarms. 

\textit{Conclusion:}
We analyzed the VM with retarded interactions, as they are expected from natural delays between sensing and reaction. It provides an alternative to the rotational inertia hypothesis for reconciling the discrepancies in the dynamical scaling and relaxation  between the standard VM and natural swarms \cite{Cavagna2018,Cavagna2018a}. While the navigation of insects and other flying species is certainly influenced both by inertia and time delay, our focus on delay could help to better understand their relation to feedback-driven robotic~\cite{mijalkov2016engineering} or micro-particle~\cite{khadka2018active,Lavergne2019,Munos2021} swarms. Especially in the latter, current experimental techniques~\cite{Franzl2021} allow inertial effects to be suppressed so that only the unavoidable time delay remains.

\begin{acknowledgements}
\textit{Acknowledgements:} We acknowledge funding through a DFG-GACR cooperation by the Deutsche
Forschungsgemeinschaft  
(DFG Project No 432421051)  and by the Czech Science Foundation (GACR Project No 20-02955J). V. H. was supported by the Humboldt Foundation. D.G. acknowledges funding by International Max Planck Research Schools (IMPRS). The time-correlation functions were computed at the GPU-Cluster Clara at the Scientific Computing Center in Leipzig. We are grateful to Lokrshi Dadhichi for his helpful remarks concerning the final manuscript. Support by Kiril Panayotov Blagoev during a preliminary stage of the project is also acknowledged.
\end{acknowledgements}

\bibliographystyle{apsrev4-2}	
\bibliography{references}	

\begin{thebibliography}{42}
\bibitem [{box()}]{boxplot}%
  \BibitemOpen
  \href@noop {} {}\bibinfo {note} {Box-plot depicts statistics on data sets.
  The central mark indicates the median, and the bottom and top edges of the
  box indicate the 25th and 75th percentiles, respectively. The whiskers extend
  to the most extreme data points not considered outliers, and the outliers are
  plotted individually using the '+' symbol. (Taken from MATLAB
  documentation.)}\BibitemShut {NoStop}%
\end{thebibliography}



\appendix

\end{document}